\documentclass[final,5p,times,twocolumn]{elsarticle} 

\usepackage{hyperref}

\usepackage{graphicx}

\usepackage{amssymb}

\usepackage{lineno}

\usepackage{rotating} 

\biboptions{comma,square}

\journal{Nucl. Instr. Meth. Phys. Res. A}

\def\znbb{0$\nu\beta\beta$}

\def\ge76{$^{76}$Ge}

\begin{document}

\begin{frontmatter}



\title{The \textsc{Majorana} Parts Tracking Database}


\author[lbnl]{N.~Abgrall} 
\author[pnnl]{E.~Aguayo} 
\author[usc,ornl]{F.T.~Avignone~III}
\author[ITEP]{A.S.~Barabash}
\author[ornl]{F.E.~Bertrand}
\author[JINR]{V.~Brudanin}
\author[duke,tunl]{M.~Busch}	
\author[usd]{D.~Byram} 
\author[sdsmt]{A.S.~Caldwell} 
\author[lbnl]{Y-D.~Chan}
\author[sdsmt]{C.D.~Christofferson} 
\author[ncsu,tunl]{D.C.~Combs} 
\author[uw]{C.~Cuesta} 
\author[uw]{J.A.~Detwiler} 
\author[uw]{P.J.~Doe}
\author[ut]{Yu.~Efremenko}
\author[JINR]{V.~Egorov}
\author[ou]{H.~Ejiri}
\author[lanl]{S.R.~Elliott}
\author[duke,tunl]{J.~Esterline} 
\author[pnnl]{J.E.~Fast}
\author[unc,tunl]{P.~Finnerty} 
\author[unc,tunl]{F.M.~Fraenkle} 
\author[ornl]{A.~Galindo-Uribarri} 
\author[unc,tunl]{G.K.~Giovanetti} 
\author[lanl]{J.~Goett} 
\author[ornl]{M.P.~Green}  
\author[uw]{J.~Gruszko} 
\author[usc]{V.E.~Guiseppe} 
\author[JINR]{K.~Gusev}
\author[alberta]{A.L.~Hallin}
\author[ou]{R.~Hazama}
\author[lbnl]{A.~Hegai} 
\author[unc,tunl]{R.~Henning}
\author[pnnl]{E.W.~Hoppe}
\author[sdsmt]{S.~Howard} 
\author[unc,tunl]{M.A.~Howe}
\author[blhill]{K.J.~Keeter}
\author[ttu]{M.F.~Kidd} 
\author[JINR]{O.~Kochetov}
\author[ITEP]{S.I.~Konovalov}
\author[pnnl]{R.T.~Kouzes}
\author[pnnl]{B.D.~LaFerriere} 
\author[uw]{J.~Diaz Leon} 
\author[ncsu,tunl]{L.E.~Leviner}
\author[sjtu]{J.C.~Loach}	
\author[unc,tunl]{J.~MacMullin} 
\author[usd]{R.D.~Martin} 
\author[tunl,unc]{S.J.~Meijer} 
\author[lbnl]{S.~Mertens} 
\author[uw]{M.L.~Miller}
\author[usc,pnnl]{L.~Mizouni} 
\author[ou]{M.~Nomachi}
\author[pnnl]{J.L.~Orrell\corref{cor1}}
\author[unc,tunl]{C.~O'Shaughnessy} 
\author[pnnl]{N.R.~Overman} 
\author[unc,tunl]{R.~Petersburg}
\author[ncsu,tunl]{D.G.~Phillips II} 
\author[lbnl]{A.W.P.~Poon}
\author[usd]{K.~Pushkin} 
\author[ornl]{D.C.~Radford}
\author[unc,tunl]{J.~Rager} 
\author[lanl]{K.~Rielage}
\author[uw]{R.G.H.~Robertson}
\author[ut,ornl]{E.~Romero-Romero} 
\author[lanl]{M.C.~Ronquest} 
\author[tunl,unc]{B.~Shanks} 
\author[ou]{T.~Shima}
\author[JINR]{M.~Shirchenko}
\author[unc,tunl]{K.J.~Snavely} 
\author[usd]{N.~Snyder} 
\author[pnnl]{A.~Soin} 
\author[sdsmt]{A.M.~Suriano} 
\author[usc]{D.~Tedeschi} 
\author[blhill,sdsmt]{J.~Thompson} 
\author[JINR]{V.~Timkin}
\author[duke,tunl]{W.~Tornow}
\author[unc,tunl]{J.E.~Trimble} 
\author[ornl]{R.L.~Varner\corref{cor2}} 
\author[ut]{S.~Vasilyev} 
\author[lbnl]{K.~Vetter\fnref{ucne}} 
\author[unc,tunl]{K.~Vorren} 
\author[ornl]{B.R.~White} 
\author[unc,tunl,ornl]{J.F.~Wilkerson} 
\author[usc]{C.~Wiseman} 
\author[lanl]{W.~Xu} 
\author[JINR]{E.~Yakushev}
\author[ncsu,tunl]{A.R.~Young}
\author[ornl]{C.-H.~Yu}
\author[ITEP]{V.~Yumatov}
\author[JINR]{I.~Zhitnikov}
\author{\newline The {\sc Majorana} Collaboration}

\fntext[ucne]{Alternate address: Department of Nuclear Engineering, University of California, Berkeley, CA, USA.}

\cortext[cor1]{Tel.: +1 509 375 1899. Contact this author for application to neutrinoless double beta decay research.}

\cortext[cor2]{Tel. +1 865 574 6521. Contact this author for software and database implementation.}

\address[alberta]{Centre for Particle Physics, University of Alberta, Edmonton, AB, Canada}
\address[blhill]{Department of Physics, Black Hills State University, Spearfish, SD, USA}
\address[ITEP]{Institute for Theoretical and Experimental Physics, Moscow, Russia}
\address[JINR]{Joint Institute for Nuclear Research, Dubna, Russia}
\address[lbnl]{Nuclear Science Division, Lawrence Berkeley National Laboratory, Berkeley, CA, USA}
\address[lanl]{Los Alamos National Laboratory, Los Alamos, NM, USA}
\address[uw]{Center for Experimental Nuclear Physics and Astrophysics, 
and Department of Physics, University of Washington, Seattle, WA, USA}
\address[unc]{Department of Physics and Astronomy, University of North Carolina, Chapel Hill, NC, USA}
\address[duke]{Department of Physics, Duke University, Durham, NC, USA}
\address[ncsu]{Department of Physics, North Carolina State University, Raleigh, NC, USA}
\address[ornl]{Oak Ridge National Laboratory, Oak Ridge, TN, USA}
\address[ou]{Research Center for Nuclear Physics and Department of Physics, Osaka University, Ibaraki, Osaka, Japan}
\address[pnnl]{Pacific Northwest National Laboratory, Richland, WA, USA}
\address[ttu]{Tennessee Tech University, Cookeville, TN, USA}
\address[sdsmt]{South Dakota School of Mines and Technology, Rapid City, SD, USA}
\address[sjtu]{Shanghai Jiao Tong University, Shanghai, China}
\address[usc]{Department of Physics and Astronomy, University of South Carolina, Columbia, SC, USA}
\address[usd]{Department of Physics, University of South Dakota, Vermillion, SD, USA} 
\address[ut]{Department of Physics and Astronomy, University of Tennessee, Knoxville, TN, USA}
\address[tunl]{Triangle Universities Nuclear Laboratory, Durham, NC, USA}


\date{\today}

\begin{abstract}
The \textsc{Majorana} \textsc{Demonstrator} is an ultra-low background physics
experiment searching for the neutrinoless double beta decay of \ge76.
The \textsc{Majorana} Parts Tracking Database is used to record
the history of components used in the construction of the \textsc{Demonstrator}.
The tracking implementation takes a novel approach based on the schema-free database technology CouchDB.
Transportation, storage, and processes undergone by parts such as machining or cleaning 
are linked to part records. Tracking parts provides a great logistics
benefit and an important quality assurance reference during construction.
In addition, the location history of parts provides an estimate
of their exposure to cosmic radiation. 
A web application for data entry
and a radiation exposure calculator have been developed as tools for
achieving the extreme radio-purity required for this rare decay search.
\end{abstract}


\end{frontmatter}




\section{Introduction}
Detecting neutrinoless double beta decay (\znbb) would prove neutrinos are their own antiparticle, demonstrate that lepton number is not conserved, and may provide insight into the mass hierarchy of the neutrino family \cite{avignone_neutrinos}. The \textsc{Majorana} \textsc{Demonstrator} \cite{AHEP2014,mj_elliott} is part of a phased approach towards a tonne-scale search for \znbb\ in \ge76, with a background goal at or below 1 cnt/(ROI-t-year) in the 4~keV wide region of interest (ROI) around the decay's 2039~keV $Q$-value. The \textsc{Demonstrator}'s expected extreme radio-purity and low energy threshold also make it sensitive to signals from dark matter candidates known as weakly interacting massive particles (WIMPs). The background goal for the \textsc{Demonstrator} is achieved through ultra-pure, low radioactivity construction materials in the immediate vicinity of the \ge76 detectors; passive shielding of naturally occurring $\gamma$-ray and neutron radiation using copper, lead, and polyethylene; active shielding against cosmic-ray generated muons using 4$\pi$-coverage plastic scintillator veto panels; and passive shielding against the majority of cosmic-ray generated particles by locating the \textsc{Demonstrator} deep underground at the 4850$^{\prime}$ level of the Sanford Underground Research Facility (SURF) \cite{SURF1,SURF2}. Furthermore, during the construction process many parts are transported or stored above ground and are thus subject to cosmogenic activation: the creation of unstable isotopes in detector components due to neutrons, protons, and muons generated by cosmic rays. These unstable cosmogenic isotopes created in the detector components may later decay and create background for the \znbb\ measurement.

Achieving the specified low background goal of the \textsc{Demonstrator} requires a multifaceted strategy including not only the detector shielding described above, but also assembly in an ultra-clean environment and careful material selection, screening, and handling. A comprehensive parts tracking effort is employed by the \textsc{Majorana} Collaboration to address the need for complete process knowledge of the low background construction materials both in terms of fabrication and above ground cosmic ray exposure. A single repository for referencing part history provides an important quality assurance check during construction. A part's history must be understood prior to integration into the production hardware. The complete \textsc{Demonstrator} is composed of thousands of individual parts and a parts tracking database thus aids in documenting the experiment's logistics associated with the fabrication and assembly of those parts. Additionally, tracking the total cosmic ray exposure enables an estimation of the cosmogenic activation and subsequent background radiation originating from detector components.

The requirements for parts, materials, and components tracking is a generic problem in many fields including industrial work, military logistics, and consumer merchandise. The parts tracking database solution described in this report is an example of a non-relational database method for addressing the part tracking challenge when the record fields defined for each component are variable and/or under development. Specifically, the ability to add record fields as needed while maintaining the ability to search the database and without having to rebuild prior records is highly advantageous for component tracking systems lacking a strong \textit{a priori} structure of record fields.

The \textsc{Majorana} Parts Tracking Database (PTDB) is driven by a CouchDB \cite{couchdb_guide,Cox201263} database and JavaScript web application. The PTDB has a web application front-end for data entry and a Python-based exposure calculator. The exposure calculator converts storage and transportation history into time and elevation profiles used to estimate material activity, which can be folded into Monte Carlo background simulations. This paper reviews the logistics of creating an ultra-low background physics experiment, the technologies used to create the user-interface web application, and how database information is stored and queried. The final sections of the paper detail the structure of the web application and exposure calculator, as well as reflect on the benefits of tracking part history for the \textsc{Demonstrator}.

\section{Parts tracking}
Part history is utilized during construction to ensure components meet quality requirements for fabrication of the \textsc{Demonstrator}. Detailed part histories also assist \textsc{Majorana} collaborators in identifying parts that may be excessively radioactive, allow for comprehensive background simulations, and aid in troubleshooting if a source of background is found within the experiment. This section presents first an overview of parts handling logistics and its implications to the database design (\ref{Logistics}) and second the software used to implement the database tracking to address logistics challenges (\ref{PTDBSoftware}).

A key to the organization of the parts tracking effort is the definition of three primary record types: \textit{parts}, \textit{assemblies}, and \textit{histories}. Parts are individual units fabricated from ``stock'' material. Assemblies describe collections of individual parts and are assigned an unique identifier. Histories are records describing steps and processes that alter and/or impact parts and assemblies. These three classes of records provide the functionality required to describe the complete fabrication and construction process of the \textsc{Demonstrator}.

\subsection{Logistics overview}\label{Logistics}
The construction logistics of the \textsc{Demonstrator} divide roughly into three stages for consideration of part tracking: material production (\ref{GeCuMaterials}), part fabrication (\ref{PartFab}), and assembly (\ref{PartAssembly}). This subsection addresses the issues associated with part tracking logistics for each stage in turn.

\subsubsection{Primary construction materials}\label{GeCuMaterials}
\begin{figure}[!t]
\centering
\includegraphics[width=\columnwidth]{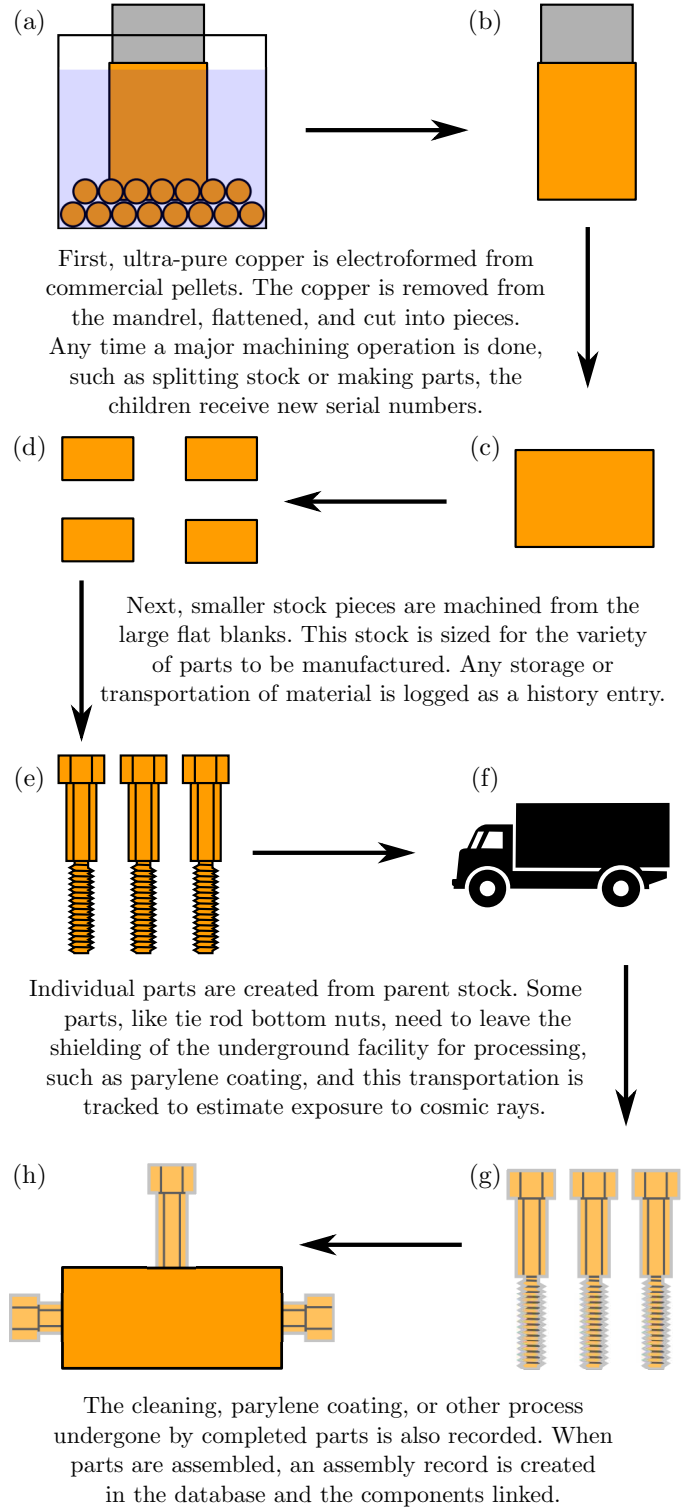} 
\caption{The life of a part, from commercial copper pellets to finished assembly, is illustrated above. Each smaller stock piece or part created from a parent stock piece receives a new serial number for tracking.
Storage, transportation, and processes such as machining or cleaning are tracked for each part and
referenced when parts are selected for use. When an assembly is created from completed parts 
the components are linked to a new assembly record in the database.}
\label{example_part}
\end{figure}

Most components in the \textsc{Demonstrator} are custom made by the \textsc{Majorana} Collaboration or its contracted vendors and are designed to be as low-mass and radio-pure as possible. In addition to the germanium crystals used as both the sources and detectors for the neutrinoless double beta decay experiment, the \textsc{Majorana} Collaboration electroforms the world's purest copper \cite{EFCu-Microscopic,EFCu-Reduction} on stainless steel mandrels in an underground electroforming facility (Figure \ref{example_part} (a)) at SURF. These two materials comprise the largest masses of materials used in the inner portions of the \textsc{Demonstrator}'s detector system.

Furthermore, of the various materials used in the inner detector system, copper and germanium are the most impacted by cosmogenic activation. As germanium is the sensitive detection media in the experiment, cosmogenic activation of the germanium crystals directly contributes background events. The cosmogenic isotope $^{68}$Ge is a particular concern as the daughter isotope $^{68}$Ga is also radioactive and can produce background events in the same energy range as those of the sought-after neutrinoless double beta decay signal of $^{76}$Ge.

The electroformed copper's value to the experiment is its ultra low concentrations of uranium and thorium, making it an excellent low-background construction and shielding material. This low background material is potentially squandered if cosmogenic activation creates elevated levels of $^{60}$Co, hence the desire to shield the electroformed copper from surface levels of cosmic rays. In the following subsections the fabrication of copper parts is used to showcase the implementation of cosmogenic activation tracking in the PTDB.

\subsubsection{Part fabrication}\label{PartFab}
To keep as much material as possible from leaving the shielding of the SURF underground, the \textsc{Majorana} Collaboration operates a clean 
machine shop at the 4850$^{\prime}$ level of SURF. The material origin and manufacturing history (parent stock tree and machining operations) 
of each part is recorded in the database, as well as any storage, transportation, or other processes a part undergoes.

To uniquely identify parts and stock used by the \textsc{Demonstrator} a serial number is laser engraved into metal and plastic parts.
Some parts are too large or small to be easily engraved, or are otherwise unsuitable for laser engraving, and these
components are identified by metal stamps or labels. Groups of small identical parts are bagged together
and a range of serial numbers is listed on the bag.

In addition to the germanium and copper used in the \textsc{Demonstrator}, steel, PTFE, and other plastics are present in smaller amounts. Within this limited palette of construction materials the parts -- once cleaned -- are visually nearly indistinguishable. This is especially a concern for distinguishing underground electroformed copper from commercial copper which is used in some circumstances and for prototyping test components. Distinguishing visually identical parts produced from different stock material is one of the vital aspects of the part tracking implementation and use.

The relationship of source materials to the manufactured parts or derived small stock materials is tracked within the PTDB.
This relationship is called ``parent-child'' in which the original material is the parent and the derived part or smaller stock is the child.
In Figure \ref{example_part}, steps (c), (d), and (e) result in ``children'' treated as new parts.
When a parent-child relationship is established the child record inherits the parent record's material type. 
The parent-child relationship is strictly one to many.  No part can have more than one parent, as the parent-child link
represents a reducing process, such as machining parts or cutting cables.

The types of history entries attached to part records include storage, transportation, machining, and a generic process entry. 
A history entry refers to one or more locations,
a list of which is stored in a separate document in the database. 
In this list, each location has address meta-data
associated with it, and the storage and transportation entries can be used to estimate exposure to cosmic rays. 
This is especially 
important if, as in step (f) of Figure \ref{example_part}, parts must be taken above-ground for some reason, a typical example being electron beam welding of joints that must hold high-vacuum, or individual parts such as \textit{tie rod bottom nuts}.

\begin{figure}[!t]
\centering
\includegraphics[width=\columnwidth]{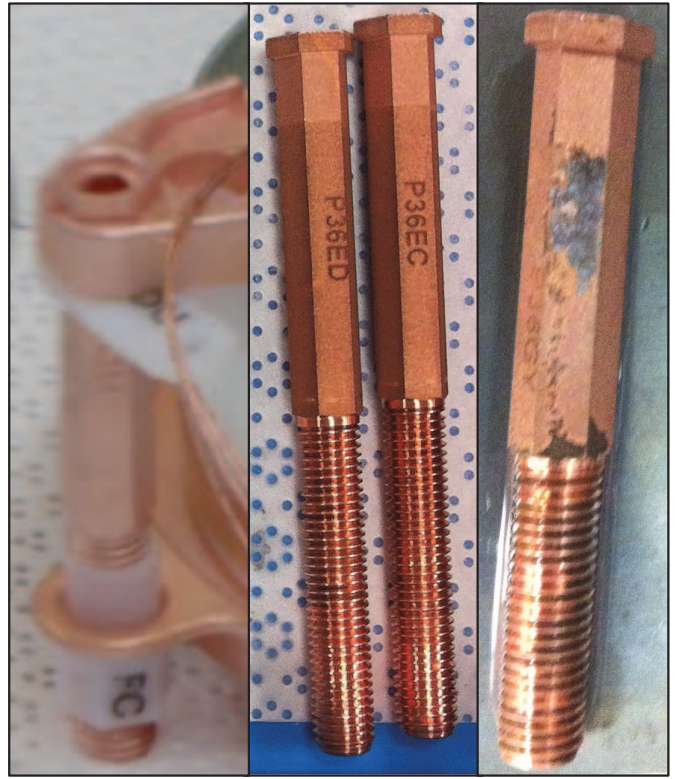}
\caption{Hollow hex rods: Part P33QU made from OFHC copper in detector unit P34Y7 (left panel), photographed through glovebox window. Parts P36ED and P36EC made from electroformed copper (center panel). Part P36GY made from electroformed copper that has had a mishap during fabrication (right panel).}
\label{HHRs}
\end{figure}

\begin{figure}[!t]
\centering
\includegraphics[width=\columnwidth]{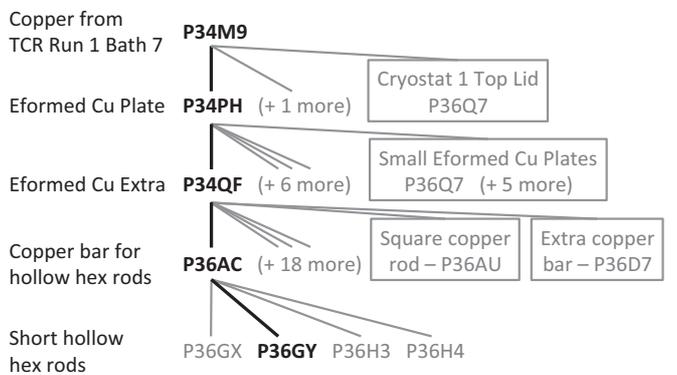}
\caption{The parents, siblings, and cousins of part P36GY as determined from the PTDB. Not all child relationships are shown for parts presented in gray.}
\label{TrackP36GY}
\end{figure}

Machining a part irreversibly changes its structure.  Many subtle details may differentiate
two otherwise very similar looking parts. Tracking the machining date, machinist, and \textsc{Majorana} drawing number allows
ambiguities to be resolved when a component's history is called into question. 
For instance, the length of a set of \textit{hollow hex rods} (custom copper bolts) is fine tuned for different \textit{detector units} that depend on the size of the housed germanium crystal, and therefore the correct length hollow hex rod must be used for assembly. Likewise distinguishing the source material for the otherwise identical looking parts is important. Figure~\ref{HHRs} shows several hollow hex rods. These hollow hex rods are made of OFHC copper (left panel) and electroformed copper. Using the serial numbers to perform a simple search reveals this fact. Perhaps more interestingly, the right most electroformed hollow hex rod shown in Figure~\ref{HHRs} has had a mishap during fabrication. While in this case the issue was only associated with this particular part, it is in principle possible that a fabrication process concern affects similarly fabricated parts. If hollow hex rod P36GY is used as a potential example case, the PTDB reveals the parent/child relationships of various parts fabricated from the same initial piece of electroformed copper produced in the underground Temporary Cleanroom (TCR) as well as any shared process history records. As Figure~\ref{TrackP36GY} shows, not only is the direct path from the initial electroformed piece (P34M9) available, so are all the branches related to the fabrication of other components. In principle if the point in the fabrication process is determined that is causing a problem, all parts down-stream from the process in question can be readily identified as each solid line connecting parts in Figure~\ref{TrackP36GY} corresponds to a \emph{single} click on a parent/child link within the PTDB user interface.

In addition to machining, components used by the \textsc{Majorana} Collaboration go through many processes. Cleaning, for instance, is a complicated task for an ultra-low
background experiment. There are many different cleaning processes a part may undergo, and a part's cleaning history
must be verified before integration into detector hardware. For example, history entries stored alongside part records in the PTDB
allow scientists to verify whether a part has been acid etched, ultrasonically cleaned, or wiped with ethanol/isopropanol.
There are many other process types and not every type can be accounted for during application development. 
As such, any user can 
easily add process types directly from the User Interface (UI). Some processes require an official \textsc{Majorana} procedure to be
followed and entry of the procedure number is required.

\subsubsection{Assembly}\label{PartAssembly}
An object which is made from more than one individual
part is known as an assembly. The PTDB provides each assembly with its own serial number and records information about it.
A part can be a member of only one assembly, although an assembly may be a member of another assembly.
If a history entry is applied to an assembly, it is also applied to each component or sub\-assembly in the assembly tree. 

The assembly types in the \textsc{Demonstrator} are taken directly from the project engineering design models and represent logical 
groupings of parts, such as \textit{detector units} (described in section \ref{PartFab}), \textit{detector strings} (a stack of 4-5 detector units), or \textit{cryostats} (a container holding 7 detector strings). Some assemblies contain multiple identical parts such as the three hollow hex rods in a detector unit.
Each component is assigned a position number which correlates directly to the simulation geometry used to model radiation transport within the \textsc{Demonstrator}.

During the assembly of detector components detailed notes are kept in physical and electronic logbooks. This information on the specifics of the assembly process is then entered into the PTDB. Since the PTDB provides a mapping of fabrication assembly history and the project engineering design models provide the radiation transport simulation geometry, the PTDB and the simulation geometry can later be merged to provide an ``as-built'' estimation of radioactive impurities on a part-by-part basis drawn from the cosmic ray exposure history and/or reports on screening assays for radioactive impurities in the construction materials. This re-merger of the PTDB information with the simulation geometry is a planned future development.

\subsubsection{Logistics summary}\label{LogisticsSummary}
Organizing and tracking information about thousands of components is a
challenge faced by many organizations, especially when production occurs in multiple and physically distant locations. 
Providing a web interface for data entry and retrieval allows a non-expert user to interact with data stored in the \textsc{Majorana} PTDB.
In this way any web browser becomes a tool that can be used to query information about a part's location or history.
This greatly increases the efficiency of quality assurance checks compared to traditional electronic log books, and the structured data can later be analyzed to calculate information such as cosmogenic activation.

\subsection{Parts tracking database implementation}\label{PTDBSoftware}
As described further in section \ref{WebAppStruct}, a JavaScript-based web application was developed by the \textsc{Majorana} Collaboration as an easy-to-use tool for data-entry.
Adopting modern programming techniques, using CouchDB, and leveraging libraries like JQuery and Backbone.js (see appendix for a complete list) allowed the web application to be easily extendable and maintainable throughout the project lifetime (see Figure \ref{simple_structure}). 
This feature of extendibility was one of several critical points for the software selection choice as the \textsc{Demonstrator} was in imminent need of a part tracking solution and had not determined a complete and comprehensive set of data entry fields required for executing a traditional strong relational database solution.
A small Python application was developed to examine  history and part records in the database and add exposure estimates to the parts histories. 
From these estimates, activation estimates will be made.
Separating the Python-based activation calculation from the database functionality was chosen to allow future improvements and developments in the activation calculation methodology -- an anticipated point of study and research interest -- without requiring modification of the database record interface, storage, retrieval, or query features.

\subsubsection{CouchDB}
CouchDB \cite{couchdb_guide} is a document database, developed to provide low-overhead, schema-less data storage and retrieval.  The fundamental entity in a CouchDB database is a document, expressed in JavaScript Object Notation (JSON) \cite{json}, identified by a single key and revision number.
Unlike Structured Query Language (SQL) databases, the database does not enforce any schema for the content of a document.  Knowledge of the internal document structure is entirely contained in user-developed code.  CouchDB does provide hooks for data queries based on optimized map-reduce functions, ``views'',  data presentation, ``lists'' and ``shows'', user access control and enhanced document update.  
The functions invoked by these hooks are written in JavaScript and stored as documents in the database.

JSON provides a human-readable
standard for data representation which is a subset of JavaScript syntax. Data are represented as object
literals that map key strings to values which may be strings, numbers, arrays, or other object literals. 
JSON parses naturally into JavaScript objects and other languages, such as Python, that have easy-to-use
JSON parsing libraries.
The choice of JSON as the underlying data representation also allows relative ease of development of future customized parsing software or application migration.

The application programming interface (API) for CouchDB is based on HyperText Transfer Protocol (HTTP) requests.
HTTP is the interface of the web and interactions with CouchDB are handled with familiar HTTP
requests such as GET, PUT, POST, and DELETE. 

Each part, assembly, and history entry is represented by a JSON document in CouchDB.  A relationship between these objects
is established by storing the serial numbers of linked parts in the appropriate JSON object.
For example, in a parent-child relationship the parent record stores the serial numbers
of its children in an array. The children all store the serial number of their parent. 
Both part and assembly documents contain a 
required information object, which contains values such as a name, record creation and last update date, and a comments.
They also hold an array for their linked history entry IDs. 
Part records have additional fields for relatives (assembly, parent, and children) as well as material type.  Assemblies have information
about the assembly type and an array for components, each entry of which details its serial number and role number. 
Membership history is also tracked
in an assembly record by serial number, including the date and whether the operation was to add or remove a component.

The structure of history entry records varies based on the type of history entry. All history records have a field for the history type,
as well as an array of linked part and assembly serial numbers. Each has a location and date field, and the transportation entries
hold a location and date for both the shipment and 
receipt of the linked records. Machining records have an additional field for the
\textsc{Majorana} drawing number used during manufacturing. Similarly, process records include a field for the procedure followed.

Updates to documents stored in CouchDB are handled through multi-version concurrency control (MVCC). Using MVCC
circumvents the needs for file locks while writing to the database. Rather than locking a database record during
a write operation, a new version of the file is is created and served by default. 
When a record is requested, the latest version is sent to the user, although older versions may be requested as well.
They are distinguished by a revision number that is required when writing a document back 
to the database. If the revision number of the document to be written does not match the revision of the current
copy in the database,
a document update conflict occurs and it is left to the application to resolve the conflict.  In the PTDB the user must
re-open the record in question and get the new version.  This is expected to be a very rare occurrence due to the relatively low concurrent usage.  If the revision numbers match,
the document is saved with a newly generated revision number.

Besides directly requesting a specific document, CouchDB takes advantage of the MapReduce paradigm for data
queries. MapReduce is an approach which optimizes a given query for parallel/cluster computing \cite{mapreduce_dean}. First a map function
is defined which is applied to all of the data in the database. It filters and sorts documents and returns key-value
pairs. For instance, a user may wish to return all database documents of some type, with a specific attribute used as
a key to access the filtered documents. 

The reduce step takes the output of the map function (sorted key-value pairs) and summarizes the result. Rather than returning a dictionary
of key-value pairs, it returns a single value. The reduce
step is optional and most queries do not require one. MapReduce is implemented in CouchDB within view documents. Each view has
separate map and reduce JavaScript functions.  Only the map function is required. 
View results are updated when the view is requested.  CouchDB optimizes this process by only reprocessing records changed since the last update, and updating related views together.  The results are saved in database records for subsequent requests, reducing view access time. \cite{couchdb_guide}.

\subsubsection{JavaScript and web applications}
As both client computers and data-serving cloud power grow, the web browser has become a
platform for application development. Web applications and services are undergoing
a drastic departure from their humble beginnings \cite{ginige_web}. 
Using a web application for the PTDB eliminates the need to have users re-install or update the application when enhancements or bug fixes are made by the database manager. Technologies proven on the web such as HyperText Markup Language (HTML)
and JavaScript provide a graphical user interface and interactivity. 

The Parts Tracking Database web portal is designed
to allow scientists to record important data and present a useful interface for inspection of the database. The software follows an 
object-oriented approach.  JavaScript does not support class-based inheritance as
in C++ (or similar languages), rather, it implements behavior reuse through prototype-based inheritance \cite{gama_js_objects}. If a desired
property is not found on an object, the property access is delegated up the prototype chain. Traditional subclassing
can be emulated by creating a new object and masking properties of the object's prototype.
In this paper the term class is used as a generic reference and may speak of
classes and prototype objects interchangeably.

Many web applications adopt what is known as a model-view-controller (MVC) decomposition
as a strategy for application structure. This approach separates the data objects from
their representation. In the web environment this is a natural strategy as view information
may deal with HTML and Cascading Style Sheets (CSS), which are entirely separate from the data's area of concern.
The MVC pattern was a natural evolution in software stemming from design by
separation of responsibility and the development of graphical user interfaces \cite{krasner1988description}. 
Many different realizations
of the concept exist under various names.
The Parts Tracking Database uses a Model-View framework known as Backbone.js. Backbone provides
base model and view objects with APIs designed for use together.   The controller function is implemented within the view object 
and a router that maps URLs directly to functions and events.

The Backbone Model 
object has a set of attributes added by the developer which are accessed through
built-in `get' and `set' functions.
The model data representing parts, assemblies, and history entries are stored in 
the CouchDB database. 
When a model object of these types is created, database JSON data is used to populate the instantiated model's
internal fields.
Data is sent back to the
database when a multi-model interaction takes place or when a document is saved from the user interface.
Multi-model interactions occur when history data is added or modified from within a part or assembly view.
When a model attribute is changed, the model triggers a change event, which is picked
up by any view objects listening to that model.

The Backbone View contains the logic and functions for interactions with the page content
and listens for model change or user input.
For instance, a model change event can cause a view to update 
the contents of a textbox.
When user input is detected (such as text entry or button clicks) the view acts 
on the detected events to update model data.
Though a view has a reference to the model object it represents,
the link may only be used for querying model state or direct data entry (for instance,
changing a selected value).

More complicated multi-model interactions, such as linking relatives or history entries, are controlled by
mediators that are singleton Backbone Collection subclasses. The Collection object provides a
flexible container for Backbone Models. Using collections as mediators
decouples models from views or other models, making the
application easier to maintain and extend. 

\subsubsection{Cosmic ray exposure calculator}
\begin{table*}[ht!] 
\renewcommand{\arraystretch}{1.2} 
\caption{FedEx Tracking information (upper left table) and the inferred set of transportation and storage record entries for the PTDB (lower table). All dates occur in the calendar year 2012. The figure shown in the upper right presents the altitude variation experienced by the part during shipment, with a notable day long stop in Denver, CO located at the red arrow. \label{tab:FedEx}}
\begin{center}
\begin{tabular}[t]{cc}
\begin{tabular}{lll}
\multicolumn{3}{l}{FedEx Tracking data (reverse chronological order)} \\ \hline
￼
￼Date, Time & Activity & Location \\ \hline
￼Jun 13, 12:06 & Delivered & Lead, SD \\
￼Jun 13, 10:37 & Out for delivery & Rapid City, SD \\
￼Jun 13, 8:50 & At local facility & Rapid City, SD \\
￼Jun 13, 5:00 & At local facility & Rapid City, SD \\
Jun 12, 21:10 & In transit & Denver, CO \\
Jun 11, 8:32 & In transit & Boise, ID \\
Jun 9, 2:50 & In transit & Hermiston, OR \\
Jun 8, 20:45 & Left FedEx origin & Pasco, WA \\
Jun 8, 14:23 & Picked up & Richland, WA \\ \hline
\end{tabular}
&
\parbox{1.05\columnwidth}{\begin{center}\includegraphics[width=1.05\columnwidth]{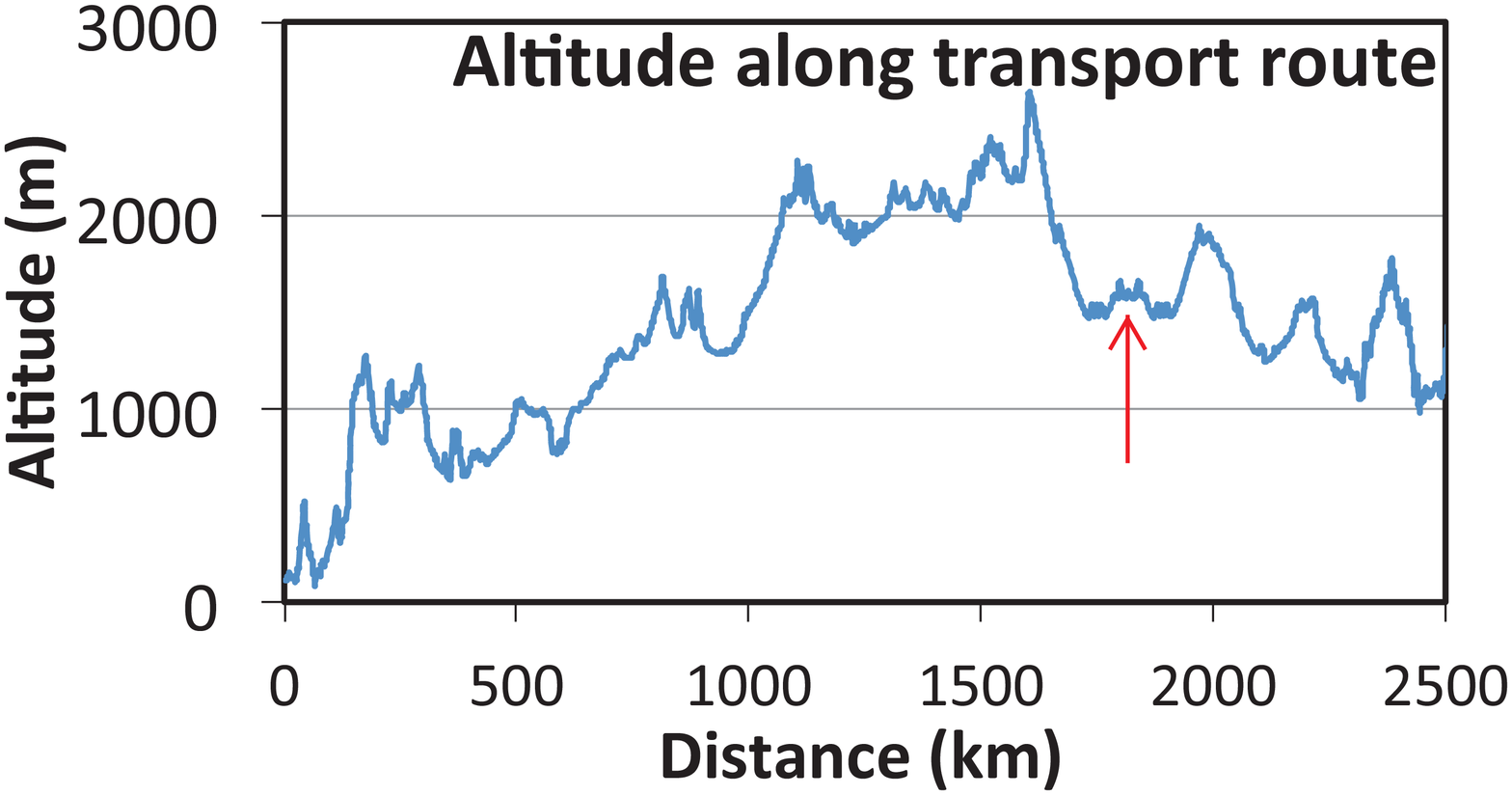}\end{center}} \\ 
\end{tabular}
\\
\begin{tabular}{c}
\\
\end{tabular}
\\
\begin{tabular}{cll|c}
\multicolumn{3}{l}{Parts Tracking Database records information (chronological order)} & \multicolumn{1}{|l}{Exposure calculation} \\ \hline
Record & Start -- End & Movement or Location & Sea-level Equivalent \\
Type & (Date, Time) -- (Date, Time) & (City, State) & Exposure (hours) \\ \hline
Transport & Jun 8, 14:23 PDT -- Jun 8, 14:43 PDT & Richland, WA $\rightarrow$ Pasco, WA & 0.4 \\
Storage & Jun 8, 14:43 PDT -- Jun 8, 20:45 PDT & Pasco, WA & 4.6 \\
Transport & Jun 8, 20:45 PDT -- Jun 8, 21:30 PDT & Pasco, WA $\rightarrow$ Hermiston, OR & 0.9 \\
Storage & Jun 8, 21:30 PDT -- Jun 9, 2:50 PDT & Hermiston, OR & 6.2 \\
Transport & Jun 9, 2:50 PDT -- Jun 9, 8:03 MDT & Hermiston, OR $\rightarrow$ Boise, ID  & 11.2 \\
Storage & Jun 9, 8:03 MDT -- Jun 11, 8:32 MDT & Boise, ID & 114.9 \\
Transport & Jun 11, 8:32 MDT -- Jun 11, 20:42 MDT & Boise, ID $\rightarrow$ Denver, CO  & 95.2 \\
Storage & Jun 11, 20:42 MDT -- Jun 12, 21:10 MDT & Denver, CO & 133.5 \\
Transport & Jun 12, 21:10 MDT -- Jun 13, 3:12 MDT & Denver, CO $\rightarrow$ Rapid City, SD & 29.8 \\
Storage & Jun 13, 3:12 MDT -- Jun 13, 10:37 MDT & Rapid City, SD & 21.1 \\
Transport & Jun 13, 10:37 MDT -- Jun 13, 11:34 MDT & Rapid City, SD $\rightarrow$ Lead, SD & 2.9 \\
Storage & Jun 13, 11:34 MDT -- Jun 13, 12:06 MDT & Lead, SD & 2.7 \\ \hline
\end{tabular}
\end{center}
\end{table*} 

In addition to the PTDB web application for data entry, a Python program was created for
examining records in the database and calculating cosmic ray exposure based on storage and 
transportation times and locations. The exposure calculator examines database records
sequentially and non-interactively inserting calculated values for exposure based on transportation and storage data.
The calculator can
also be run interactively in server mode, using the Twisted library, and it accepts HTTP requests for
part exposure data. This feature is potentially useful when the cosmogenic activation calculation is finalized, likely well after completion construction of the \textsc{Demonstrator}.
The locations record in the database contains addresses which
are converted to latitude, longitude, and elevation, ``geocode'',  using the Google Maps API. 
Elevation and time data can
then be used to calculate expected total cosmic-ray exposure for a given transportation route
or storage location.

Twisted is an event-driven network layer which can use many networking protocols.
The HTTP functionality is used to return exposure data as JSON strings. By
using Twisted, the application obtains multi-threaded concurrency, allowing
the exposure calculator to take advantage of a multi-core processor when running in 
server mode. Using the framework simply involves writing callback functions for the appropriate request
Universal Resource Identifier (URI).

Because more than one part may be linked to a history event, the calculated exposure is
stored in the storage/transportation records, along with a calculation timestamp, or a flag
if something has gone wrong during calculation. When not operating in server mode, the Python
exposure calculator examines the history records of the database and calculates exposure for
new history entries or if a particular entry's timestamp is out of date. 

Exposure data are also added to a part's database record, using associated history record
data. If a new history record is added (or the timestamp is out of date) the part
exposure will be updated on the next evaluation. The Google Maps API  may return multiple possible
routes between two locations and the correct route must be chosen or created
manually for an accurate estimate of exposure to cosmic radiation.   The software will warn users of this 
ambiguity and enable later selection of the appropriate route.

An example of the usage of the cosmic ray exposure calculator is provided by the shipment of a mandrel of electroformed copper grown in PNNL's shallow underground laboratory~\cite{PNNLUG} and transported by FedEx Ground Freight to MJD's underground laboratory at SURF. On June 8, 2012 part P34F4 was picked-up in Richland, WA, USA The route taken, as reported by FedEx Tracking, included passing through Pasco, WA $\rightarrow$ Hermiston, OR $\rightarrow$ Boise, ID $\rightarrow$ Denver, CO $\rightarrow$ Rapid City, SD (all within the continental USA) before arriving in Lead, SD. Table~\ref{tab:FedEx} shows the FedEx Tracking information (upper left) and how that information is transformed into transportation and storage records for use in the PTDB (lower half of Table~\ref{tab:FedEx}). The FedEx Tracking information shows the shipment took 4 days 19 hours 43 minutes. The routing information provided by the Google Maps API also gives expected travel times between FedEx facility locations and -- assuming no stops -- would required only 23 hours 18 minutes of driving to cover the 1,578 miles traveled. It is clear that understanding the duration a shipment spends ``in storage'' at different FedEx Freight facilities is key to estimating the total sea level effective cosmic ray exposure, especially considering the altitude of the Denver, CO FedEx Freight facility.

For the example part P34F4, the cosmic ray exposure code implemented an exponential parameterization of the equivalent sea-level cosmic ray exposure level as a function of altitude given by
\begin{equation}
\textrm{Exposure factor} = 5^{~\textrm{Alt.}\textrm{(feet)}~\textrm{/}~\textrm{5000}\textrm{(feet)}}
\end{equation}
as derived from more detailed cosmic ray flux models~\cite{ExposureFactor}. The equation is used to directly scale the duration spent at a given altitude to an equivalent duration of time spent at sea level. This resulted in a total sea-level equivalent cosmic ray exposure duration of 17.6 days for part P34F4 during its roughly 5 day long  shipment from PNNL to SURF. For context, the \textsc{Majorana} Collaboration has determined the activation of electroformed copper may range from 1-9 months of sea level exposure depending on the location within the experiment of the subsequently fabricated copper parts. As electroformed copper is also produced underground at SURF, the PTDB also assists in distinguishing the source of electroformed copper. Electroformed copper such as part P34F4 is used to fabricate components that have the least impact on the experimental background.

When transferring the FedEx Tracking information to the PTDB, one must reconstruct a shipping scenario based on the FedEx Tracking information combined with assumptions about the duration of the travel times and periods of storage. The lower half of Table~\ref{tab:FedEx} shows the complete set of transportation and storage records for the PTDB based upon the assumption that each transportation duration is the time reported by the Google Maps API required to drive along that route of the shipment. This method of constructing the PTDB transportation and storage records assigns any additional time between the FedEx Tracking ``scan'' times as storage time at the arrival location. Effectively this means a FedEx Tracking record listed as ``In transit'' is taken to signify the time at which the shipment is loaded and beginning the next leg of the shipping route.

The above described method of constructing the transportation and storage records for the PTDB necessarily builds-in some uncertainty associated with the time spent in FedEx Freight facilities, effectively under a ``storage'' condition. Examining the results presented in the lower half of Table~\ref{tab:FedEx} shows the sea-level equivalent cosmic ray exposure is predominately accumulated while in storage at these FedEx facilities. Specifically, for part P34F4 it is estimated that 11.8 days of sea-level equivalent cosmic ray exposure is accumulated in the FedEx Freight facilities while only 5.8 days sea-level equivalent cosmic ray exposure is accumulated during physical transit. These observations suggest adding a GPS recorder to shipments to provide a more readily understandable location history or even coupling the PTDB and cosmic ray exposure estimates with methods for assessing the cosmic ray flux along the route~\cite{muWitness}.

\section{Web application structure}\label{WebAppStruct}

\begin{figure}[!t]
\centering
\includegraphics[width=\columnwidth]{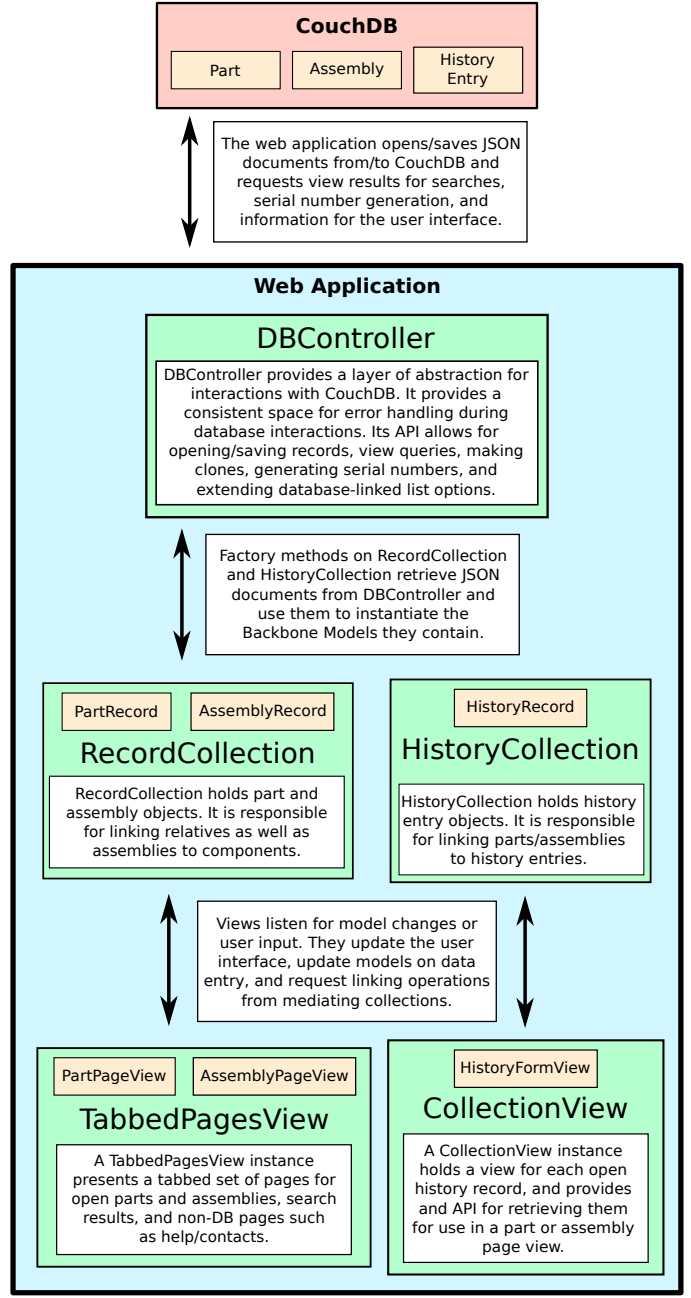} 
\caption{The above illustration demonstrates key relationships between elements of the PTDB web application
and CouchDB. JSON documents representing parts, assemblies, and history entries are stored in the database.
A DBController instance provides an abstraction layer between CouchDB and Backbone models and routers in the application.
PartRecord, AssemblyRecord, and HistoryRecord models are stored in mediating collections (RecordCollection and HistoryCollection
singletons) which carry out multi-model interactions. Views monitor individual models as well as RecordCollection and HistoryCollection.
They refresh themselves if their models trigger change events and create new views when add events are triggered
on the collections. They also listen for user-input and update models or message mediators as needed.}
\label{simple_structure}
\end{figure}

The data-entry web application is written completely in JavaScript, and is served from the same CouchDB instance
which holds the data. An object-oriented style is used with a Model-View decomposition provided
by the Backbone.js framework. Parts, assemblies, and history records are represented by models. They are contained
in two singleton collections (see Figure \ref{simple_structure}) which act as mediators. Messages passed to 
the mediators from views or models trigger inter-model linking such as adding a history entry to a part.
The view objects listen to events from and update both the models and the user interface. Application level
views listen for events from the mediating collections and create or destroy record views as needed. 

Database interactions are abstracted by a singleton controller, enhancing error handling for database requests
and providing functionality such as serial number generation and cloning. Other controllers handle data validation,
Backbone router-based web-page navigation
and data caching. The form-based data entry style of the application is made richer with a
set of form models, views, and collections. They provide interactivity beyond standard HTML forms and
specify an API for generating JSON from the data they contain.

\subsection{Models}
The real-life items of interest in the Parts Tracking Database are the parts and assemblies which compose
the experiment's hardware. These items are represented in the web application by the DBRecord model type, which is
the parent of PartRecord and AssemblyRecord. The attributes of these models point to FormCollection instances, which hold
Form model objects.

An event that a part may undergo, such as transportation, storage, or machining,  is represented by a 
HistoryRecord object. Similar to a DBRecord, a HistoryRecord has a ``collection'' attribute which points to a
FormCollection for the particular history event type. DBRecords and HistoryRecords have APIs for querying
of which assembly a PartRecord or AssemblyRecord is a member,  or rebuilding
broken relationships, in which, for example, a child points to a parent but the parent has no reference to the child, caused by a 
failure of the browser.

Because one goal of the PTDB is to present a rich user interface for data entry, the forms
themselves have model and view representations. 
Form models contain the data and logic for presenting a rich user interface, but do not contain HTML/CSS or
methods which manipulate the page content.  These exist in the FormView class tree.
The simplest type of form is a textbox and label pair. The basic Form model class stores data 
about the object -- its current value, the label and tooltip text,
and a boolean value representing whether or not the model data was validated successfully.

More complex forms include dropdown boxes containing options pulled from CouchDB. The model
stores a reference to the correct type of options for the dropdown list, for instance, whether it is a locations, people,
or process type dropdown. In addition, if the 
desired option is not present, the form can be switched into a textbox. The value entered by
the user is sanitized (non-alphanumeric characters are removed) and a key is generated based on the sanitized value. The new key-value
pair is then written to the database and is present for future use.

This database-active dropdown object is exploited for lists of locations, processes, or collaboration members. No administrative intervention
is required for the addition of information to these lists. To improve performance,
the options lists are cached in an object which specifies a short-term timeout.
If many records are opened at once and they contain multiple DB-active dropdown lists, they 
pull data from the cached list object rather than sending out HTTP requests and awaiting 
a response.

Another type of form used is a list of records, the RecordList and its children RelativesList and HistoryRecordList. 
In the CouchDB JSON record, lists of relatives or linked histories are simply arrays of database IDs.
RecordList encapsulates a collection of record values (names, dates, etc.)
as well as methods for adding/removing records from the list. Data for the collection are not loaded
until user interaction with the view to improve performance when multiple record models are opened.

Querying data from the user interface creates a SearchPage instance, which holds the results of a CouchDB view request.
As mentioned previously, queries to CouchDB are handled through views that are implementations of
MapReduce functions that operate on the data. The PTDB presents a few different search types to users including
part serial number, name, comments, linked history records, material type, or machining operations.
Partial key matching is supported for the drawing number search.
Data returned from a CouchDB view are used to instantiate a SearchResults collection which implements a variety of
sorting features for the displayed results. 

Besides directly manipulating a single record, a user may also open multiple records at once.
Controlled by the MultipleSelect model, users may add a history entry to a
large number of parts in one operation and avoid separately modifying each part. This interactivity is available from
the view displaying search results.

A user may also wish to copy data present in an existing part into one or more new parts. This
feature is called cloning and is implemented through a general operation which accepts the number
of clones to be made. Any data and history entries present in the original part are present in the
clones produced, and history entries are re-linked to the new clones such that the clone serial
numbers appear in the linked records data of the history entries.  Cloning is meant to make
data entry during parts production easier.  To prevent errors in inherited history records, assembly members may not be cloned. 

\subsection{Collections}
\begin{figure*}[!t]
\centering
\includegraphics[width=2\columnwidth]{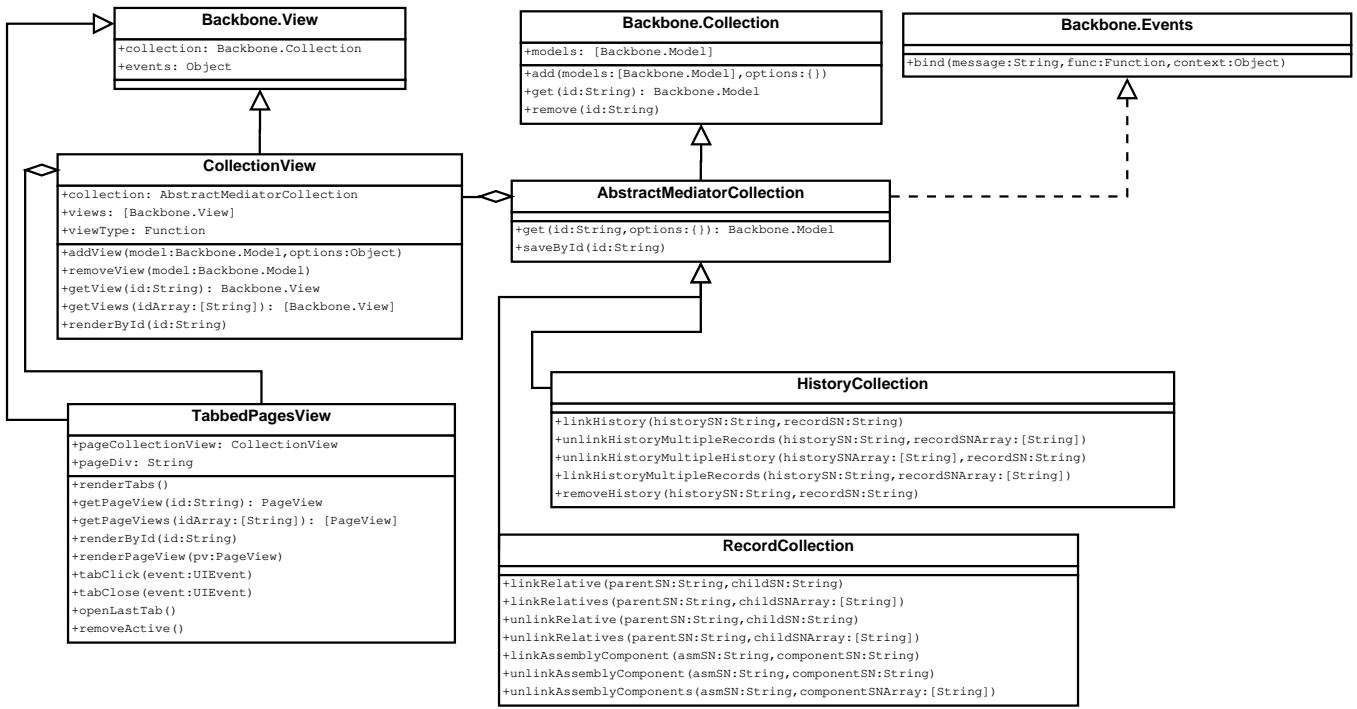} 
\caption{A structure diagram of important Collection singletons and their associated views. CollectionViews aggregate views for models in
AbstractMediatorCollections. TabbedPagesView presents the tabbed page interface which is the primary point of interaction with users, and it
does so by manipulating an internal CollectionView. HistoryCollection and RecordCollection inherit properties from AbstractMediatorCollection
and act as mediators, controlling the logic that links relatives, assembly components, and history entries.}
\label{singletons}
\end{figure*}

The tracking logic (linking relatives, assembly components, and history entries) is controlled
by singletons that subclass the AbstractMediatorCollection class (see Figure \ref{singletons}). This class is a Backbone.Collection
that has been extended with the Backbone.Events API, and acts as a mediator between DBRecord instances (parts and assemblies) and
HistoryRecord instances. The singletons are notified by events from models or views and carry out the needed
linking operations.

The AbstractMediatorCollection class also extends the
Backbone.Collection.get API for ease of use. An option can be passed to the 'get' method which
specifies whether or not to open a record from the database if it is not currently a member
of the collection.
A factory method on the collection examines the JSON
fetched from CouchDB and creates an instance of the correct type, adding the new model
to the collection. 

As mentioned before, DBRecord and HistoryRecord objects have attributes which point to FormCollection instances (subclasses
of Backbone.Collection).
FormCollections contain a set of Form models as well as some basic logic. Instances can return an array of form IDs
and specify an API for returning JSON from Form models contained in the collection. Subclasses 
group sets of related forms together and implement extra logic related to the set of forms. For instance, AssemblyForms (another Backbone.Collection
subclass)
has a method for returning the serial numbers of components in the assembly, as it contains a Form model which holds
that information. PartForms has a method for getting or setting the part's material type, grabbing the value from the
desired Form model.

\subsection{Views}
The views have access to the Backbone.Events API and listen for changes in model attributes. If a change is detected,
the view is re-rendered. Interaction with elements in the user interface triggers events which are caught by the view and the appropriate
functions are run. To modify DBRecord and/or HistoryRecord models, views pass messages to 
RecordCollection or HistoryCollection which manage linking interactions between models.

The CollectionView class holds views for models stored in AbstractMediatorCollections. HistoryFormViews
are held by a CollectionView, for instance. A tabbed interface is presented by TabbedPagesView,
which holds a CollectionView for part and assembly record views. Views which do not represent objects in the database, like those
for help or contact tabs, are also contained in TabbedPagesView. The history CollectionView and TabbedPagesView
listen for ``add'' events on their respective collections and create views for the added models as need.

As the Form models are designed around a certain user experience, there is a set of views with parent class
FormView that mirrors the Form model inheritance 
tree. There is no such expectation for DBRecord models and
in principle multiple views could represent a particular part or assembly, though currently the only representation is
through the TabbedPagesView interface.

Though history records are shared among parts, one view per open history model is used to represent each history
record. The view for a particular history entry is shared among various part/assembly views. This ensures the absence of
stale views/information in the application and reduces memory needs and view updates.

\subsection{Other objects}
Input data validation is handled by a singleton class (Validator) and the Backbone.Model validate API. The validate
method of model objects delegates to the appropriate Validator method.
A regular expression test is done and the result returned. The Validator singleton
allows for all regular expressions needed for validation to be present in a single object, reducing scattered
regular expression patterns in the application. Examples of validated fields include serial numbers, drawing numbers, and \textsc{Majorana} procedure numbers.
Invalid data triggers view notification 
and can, for instance, turn backgrounds 
red to notify the user that something has gone wrong. Invalid data can prevent record saving, or relative/history linking.

Database interactions are abstracted with the DBController singleton. It allows for error-handling during saving or 
opening documents, and requesting view results, to be primarily in a single object. It also has methods for generating new part, assembly, or history
entry serial numbers, as well as for cloning and adding to DB-active lists.

DBListController is a singleton that caches options lists stored in CouchDB, such as people, locations, or process types. Its
API allows for retrieving list options, forcing a cache reset, and checking to see if the cache is valid. The cache timeout is 
one minute by default, longer than the execution time for most tasks, but short enough to keep the application current
with changes from a user in another location. Caching the list options improves interaction speeds when many parts must be opened for manipulation. For
instance, adding a history entry to a detector string assembly may involve changes in over one hundred parts and each part contains many
DB-active lists to be fetched, and may contain history entries with more.

The MJDBRouter object extends Backbone.Router to provide address-bar controlled features. This allows basic application state
to be shared via a simple link, which might for instance open a certain record in the tabbed interface. Other use cases include
updating the address-bar when searching the database so the view type and keys are visible to the user. In this way the MJDBRouter
acts on a `message', that is the route navigated to, causing function execution with optional arguments, while displaying the route
to the user in the navigation bar.

\section{Conclusion}

Neutrinoless double beta decay, if it exists, is an exceedingly rare process.
To make a measurement or set a limit on \znbb\ or on dark matter interactions 
requires the  \textsc{Majorana} \textsc{Demonstrator} to reach very low background levels\footnote{For neutrinoless double beta decay the background rate goal is 1~cnt/(ROI-t-year) in the 4~keV wide region of interest (ROI) around the decay's 2039~keV $Q$-value. For dark matter analysis, a background rate below 1~cnt/(keV-kg-day) in an analysis region below $\sim$5~keV electron equivalent recoil energy is desired.}. A parts tracking effort which allows material type,
storage and location history, and other processes undergone to be recorded and
accessed by users or computer programs has proven to be an effective tool in a comprehensive
radio-purity campaign. In addition, tracking detector components presents a significant
logistics challenge in which the Parts Tracking Database aids.

A web application front-end for data entry allows users unfamiliar with the database implementation software to input
or query data on part type, material, location, or history. The Python cosmic ray exposure
calculator is an example of one way this data can be used to generate an estimate
of part-by-part cosmogenic radioactivity. These tools have been developed with ease of maintainability and
extensibility in mind. Backbone.js provides structure and a Model-View
decomposition to the web front-end. Techniques such as message passing through mediators
decouple much of the inter-model (linking relatives, assembly components, and history entries)
and model-view logic,  simplifying the code.

CouchDB provides a flexible data storage solution and speaks HTTP -- a natural
fit for a web application. 
Both features were deemed critical to the success of the PTDB development effort. 
The flexibility was required as the database development was to occur \emph{during} initial parts production 
and the entirety of the database structure was not known at the time of development. Access via an HTTP web-interface 
was seen as crucial to ensuring adoption of the PTDB by the \textsc{Majorana} Collaboration members 
who need to track parts, but were otherwise not focused on PTDB software development.
Libraries exist to work with CouchDB in many languages and
features such as MapReduce and Multi-Version Concurrency Control make it a powerful
alternative to the traditional SQL based database solution.
The user-base for the PTDB are the members of the \textsc{Majorana} Collaboration which consists of 
researchers from across North America and overseas whose primary desires for performance of the PTDB are 
focused on ease of data entry, query (search), and ability to easily link parts and part histories.

Low-background physics experiments can benefit greatly from comprehensive
parts and materials tracking efforts like \textsc{Majorana}'s Parts Tracking Database.
The Parts Tracking Database is a 
key component of the \textsc{Demonstrator}'s radio-purity campaign and will help the collaboration achieve
its goals of probing ultra-rare processes.

\section*{Acknowledgment}
This material is based upon work supported by the U.S. Department of Energy Office of Science, Office of Nuclear Physics under Award Numbers DE-AC02-05CH11231, DE-AC52-06NA25396, DE-FG02-97ER41041, DE-FG02-97ER41033, DE-FG02-97ER41041, DE-FG02-97ER41042, DE-SC0012612, DE-FG02-10ER41715, and DE-FG02-97ER41020. We acknowledge support from the Particle and Nuclear Astrophysics Program of the National Science Foundation through grant numbers PHY-0919270, PHY-1003940, 0855314, PHY-1202950, MRI 0923142 and 1003399. We acknowledge support from the Russian Foundation for Basic Research, Grant No. 12-02-12112. This research used resources of the Oak Ridge Leadership Computing Facility, which is a DOE Office of Science User Facility supported under Contract DE-AC05-00OR22725. This research used resources of the National Energy Research Scientific Computing Center, a DOE Office of Science User Facility supported under Contract No. DE-AC02-05CH11231.  We acknowledge the support from the U.S. Department of Energy through the LANL/LDRD Program. We thank M.~J.~Giardinelli (Pacific Northwest National Laboratory) for review of an early draft of this paper. We thank our hosts and colleagues at the Sanford Underground Research Facility for their support.

\appendix

\section{Software libraries used in the PTDB}

\subsection{Web application}
\begin{itemize}

  \item \textbf{JQuery} -- \url{http://jquery.com/}

   JQuery is a popular JavaScript library which simplifies 
   interaction with the browser Domain Object Model, 
   AJAX  (Asynchronous JavaScript and XML) requests, and simple user interface UI effects.
   The JQuery UI toolkit is also used for the datepicker
   calendar functionality.

  \item \textbf{Backbone.js} -- \url{http://backbonejs.org/}

   Backbone.js provides a model-view decomposition by providing
   Model, View, and Collection prototype objects. It also provides
   a Sync object for database interactions as well as a Router object
   for URL fragment navigation and page history. 

  \item \textbf{Underscore.js} -- \url{http://underscorejs.org/}

   Underscore is a JavaScript utility library used for common
   tasks and lending a functional programming style where
   desired.
   It is required by Backbone.js.

\end{itemize}

\subsection{Exposure calculator}
\begin{itemize}

  \item \textbf{Google Maps API v3} -- \url{https://developers.google.com/maps/documentation/javascript/}

  Google Maps API is a service that provides the route and distance between
  locations as well as the elevation via an HTTP request. The Maps API provides
  this information via the DirectionsService and the ElevationService for
  JavaScript applications.

  \item \textbf{Twisted} -- \url{http://twistedmatrix.com/trac/}

   Twisted is a Python network programming layer we use to serve
   JSON exposure data via HTTP. It is an event-driven library which
   turns URI requests into function execution, and returns data
   to the requester.

\end{itemize}


\bibliographystyle{elsarticle-num}
\bibliography{mjdptdb}






\end{document}